\documentstyle[12pt,epsfig]{article}
\def\fun#1#2{\lower3.6pt\vbox{\baselineskip0pt\lineskip.9pt
  \ialign{$\mathsurround=0pt#1\hfil##\hfil$\crcr#2\crcr\sim\crcr}}}
\newcommand{\dd}{\mbox{d}}

\newcommand{\ii}{\mbox{i}}

\newcommand{\vecc}[1]{\mbox{\boldmath $#1$}}
\newcommand{\Li}{\mbox{Li}_2}
\newcommand{\matrm}[1]{\mbox{\scriptsize #1}}

\title { Radiative large--angle Bhabha scattering in collinear kinematics }

\author {V.~Antonelli$^a$,
E.A.~Kuraev$^b$, B.G.~Shaikhatdenov$^b$
\thanks{on leave of absence from the Institute of Physics and Technology, Almaty}
\\[4mm]
$^a$ {\small\sl Dipartimento di Fisica, Universit\`a di Milano-Bicocca},\\
{\small\sl and INFN Sezione di Milano, Milano, Italia}\\
$^b$ {\small\sl Joint Institute for Nuclear Research, Dubna, 141980, Russia}\\
}

\date{}

\begin{document}
\maketitle

\begin{abstract}
The process of large--angle high energy electron--positron
scattering with emission of
one hard photon almost collinear to one of the charged particles momenta is
considered.
The differential cross section with
radiative corrections due to emission of virtual and soft real photons
calculated to a power accuracy is presented.
Emission of two hard photons and total expressions for radiative
correction are given in leading logarithmical approximation.
The latter are illustrated by numeric estimates.
A relation of the results with structure function formalism is discussed.
\\[.2cm] \noindent
\footnote{Preprint Bicocca-FT-99-13 and hep-ph/9905331}
\end{abstract}

\section{Introduction}

The process of electron--positron scattering is commonly used
for luminosity measurements at $e^+e^-$ colliders. It has almost pure
electrodynamical nature and could therefore be described to any
desired precision within a framework of perturbative QED. Nevertheless
the accuracy of modern experiments is ahead of that provided by
theory. A lot of work has recently been done to uplift the theoretical
uncertainty
to about one per mille under conditions of small--angle Bhabha scattering
at LEP1~\cite{LEP1} and afterwards up to $0.05-0.06\%$\cite{WM}. 

The large--angle kinematics of Bhabha
scattering process is extensively used for
calibration purposes at $e^+e^-$ colliders of moderately high energies,
such as $\phi$, $J/\psi$, $B$, and $c/\tau$ factories and LEP2.
At the Born and one--loop levels the process was investigated
in detail in~\cite{l1,l2,l3,l4,l5}, taking into account
both QED and electroweak effects.

In paper~\cite{Fedot}
we considered Bhabha scattering to ${\cal O}(\alpha)$ order exactly
improved by the structure function method. The latter, based on
the renormalization group approach, allows to evaluate
the leading radiative corrections to higher orders, including
all the terms $\sim (\alpha L_s)^n$, $n=2,3,\dots$,
where $L_s=\ln(s/m^2)$ is a large logarithm,
$s$ is the total center-of-mass (cms) energy of incoming particles squared
and $m$ is the mass of fermion.

To reach the one per mille accuracy it is required
to take into account radiative corrections~(RC) up to third order
within the leading logarithmic approximation~(LLA) and up to
second order in the next--to--leading approximation~(NLA).
In a series of papers several sources of these corrections were
considered in detail~\cite{2h,pairs,all,viol}.

In a recent publication~\cite{viol} the contribution due to
virtual and soft photon corrections to large--angle radiative
Bhabha scattering was calculated for the general case of hard photon
emission at large angle with respect to all charged particles
momenta. In the present work we are going to consider the complementary
specific kinematics, in which the photon moves
within a narrow cone of small opening angle $\theta_0\ll 1$
together with one of the incoming or outgoing charged particles.
Thus, the result obtained here may be used in experiments with the tagging
of scattered electron (positron) in detectors of small aperture
$\theta_0\ll 1$.

Our paper is organized as follows. In Sec.~2 the Born level
cross section of radiative Bhabha scattering is revised
in the collinear kinematics of photon emission along initial
(scattered) electron. We introduce here the physical gauge of real
photon that is extensively used in the next sections.
In Sec.~3 a set of crossing transformations which enables us
to consider in some detail only the scattering type amplitudes of loop
corrections to the process is described. Besides, we restrict
ourselves to the kinematics of hard photon emission along initial electron.
In Sec.~4 the corrections due to virtual and soft real photon emission
in the case $\vecc{k}_1\parallel\vecc{p}_1$ are considered.
The general expression for correction
in the case of hard photon emission along scattered electron is given
in Sec.~5. In Sec.~6 we consider a contribution (in~LLA) coming from
two hard photon emission and derive a general expression for radiative
correction. In conclusion we discuss the relation with structure
function approach and the accuracy of the results obtained.
Some useful expressions for loop integrals are given in the Appendix
and the results of numeric estimates are given in graphs.

\section{Born expressions in collinear kinematics}

Let us begin revising the radiative Bhabha scattering process
\begin{eqnarray} \label{proc}
e^-(p_1)\ +\ e^+(p_2)\ \to\  e^-(p_1')\ +\ e^+(p_2')\
+\ \gamma(k_1)
\end{eqnarray}
at the tree level.
We define the collinear kinematical domains as those
in which the hard photon is emitted
close (within a narrow cone with opening angle $\theta_0\ll 1$)
to the incident
($\theta_{1(2)}=\widehat{\vecc{p}_{1(2)}\vecc{k}_1}<\theta_0$)
or the outgoing electron (positron)
($\theta_{1(2)}'=\widehat{\vecc{p}'_{1(2)}\vecc{k}_1}<\theta_0$)
direction of motion. Because of the symmetry between electron
and positron we may restrict ourselves to a consideration of only two
collinear regions, which correspond to the emission of the photon along
the electron momenta. The two remaining  contributions to the differential cross
section of the process~(\ref{proc}) can be obtained
by the substitution ${\cal Q}$
\begin{eqnarray} \label{pr}
\dd\sigma_{\matrm{coll}}
&=&\left[1+{\cal Q}
\left(\begin{array}{c}p_1\leftrightarrow p_2\\p_1'\leftrightarrow p_2'
\end{array}\right)\right]
\biggl\{\dd\sigma^\gamma(\vecc{k}_1\parallel \vecc{p}_1)
+\dd\sigma^\gamma(\vecc{k}_1\parallel \vecc{p}_1')\biggr\}.
\end{eqnarray}
To begin with, let us recall the known expression~\cite{Ber73}
in Born approximation
for the general kinematics, i.e. assuming all the squares of the momenta
transferred among fermions to be large compared to the electron mass squared:
\begin{eqnarray}\label{born}
\dd\sigma_0^\gamma
&=&\frac{\alpha^3}{8\pi^2s}T\dd\Gamma, \qquad
\dd\Gamma = \frac{\dd^3\vecc{p}_1'\dd^3\vecc{p}_2'\dd^3\vecc{k}_1}
{\varepsilon_1'\varepsilon_2'\omega_1}
\delta^4(p_1+p_2-p_1'-p_2'-k_1), \\ \nonumber
T&=&\frac{S}{tt_1ss_1}\left[ss_1(s^2+s_1^2)+tt_1(t^2+t_1^2)
+uu_1(u^2+u^2_1)\right] \\ \nonumber
&-&\frac{16m^2}{\chi_2^{'2}}\left(\frac{s}{t_1}+\frac{t_1}{s}+1\right)^2
-\frac{16m^2}{\chi_1^{'2}}\left(\frac{s}{t}+\frac{t}{s}+1\right)^2
-\frac{16m^2}{\chi_2^2}\left(\frac{s_1}{t_1}+\frac{t_1}{s_1}+1\right)^2 \\
\nonumber
&-&\frac{16m^2}{\chi_1^2}\left(\frac{s_1}{t}+\frac{t}{s_1}+1\right)^2, \\
\nonumber
S&=&
4\left[\frac{s}{\chi_1\chi_2}
+\frac{s_1}{\chi_1'\chi_2'}-\frac{t_1}{\chi_1\chi_1'}
-\frac{t}{\chi_2\chi_2'}+\frac{u_1}{\chi_2\chi_1'}+\frac{u}{\chi_1\chi_2'}
\right],\\ \nonumber
&&s=(p_1+p_2)^2,\quad s_1=(p_1'+p_2')^2,\quad t=(p_2-p_2')^2,
\quad t_1=(p_1-p_1')^2, \\ \nonumber
&&u=(p_1-p_2')^2,\quad u_1=(p_2-p_1')^2,\quad \chi_i=2p_ik_1,
\quad \chi_{1,2}'=2p_{1,2}'k_1.
\end{eqnarray}
In the collinear kinematical domain
in which $\vecc{k}_1\parallel \vecc{p}_1$
the above formula takes the form
\begin{eqnarray}\label{s0}
\dd\sigma_0^\gamma(\vecc{k}_1\parallel \vecc{p}_1)&=&\frac{\alpha^3}{\pi^2s}
\frac{\dd^3\vecc{k}_1}{\omega_1}\frac{1}{\chi_1}\Upsilon F
\frac{\dd^3\vecc{p}_1'\dd^3\vecc{p}_2'}{\varepsilon_1'\varepsilon_2'}
\delta^4((1-x)p_1+p_2-p_1'-p_2') \\ \nonumber
&=&\dd W_{p_1}\dd\sigma_0((1-x)p_1,p_2), \\ \nonumber
\Upsilon&=&\frac{1+(1-x)^2}{x(1-x)}-\frac{2m^2}{\chi_1},
\qquad F=\left(\frac{s_1}{t}+\frac{t}{s_1}+1\right)^2,
\end{eqnarray}
where
\begin{eqnarray}
&&s_1=s(1-x),\quad
y_1=\frac{\varepsilon_1'}{\varepsilon}=2\frac{1-x}{a},
\quad y_2=\frac{\varepsilon_2'}{\varepsilon}=\frac{2-2x+x^2+cx(2-x)}{a},\nonumber\\
&&a=2-x+cx,\quad \omega_1=\varepsilon x,\quad s=4\varepsilon^2,\quad
\chi_1=\frac{s}{2}x(1-c_1\beta),\quad \beta=\sqrt{1-\frac{m^2}{\varepsilon^2}},
\nonumber \\
&&t=t_1(1-x)=-s\frac{(1-x)^2(1-c)}{a},\quad
c=\cos(\widehat{\vecc{p}_1\vecc{p}_1'}), \quad
c_1=\cos(\widehat{\vecc{p}_1\vecc{k}_1}), \nonumber \\
&&\dd W_{p_1}=\frac{\alpha}{2\pi^2}\frac{1-x}{\chi_1}\Upsilon
\frac{\dd^3\vecc{k}_1}{\omega_1}.
\end{eqnarray}
Here $y_i$ are the energy fractions of the scattered leptons
and $\dd\sigma_0(p_1(1-x),p_2)$ is the cross section of the elastic Bhabha
scattering process.

Throughout the paper we use the following relations among invariants
\begin{eqnarray*}
s_1+t+u_1=4m^2-\chi_1\approx 0,\qquad s+t_1+u=4m^2+\chi_1\approx 0.
\end{eqnarray*}

In the case $\vecc{k}_1\parallel \vecc{p}_1'$ we have
\begin{eqnarray}
\dd\sigma_0^\gamma(\vecc{k}_1\parallel \vecc{p}_1')&=&\frac{\alpha}{2\pi^2}
\frac{1}{\chi_1'}\tilde\Upsilon
\frac{\dd^3\vecc{k}_1}{\omega_1}(1-x)\dd\sigma_0(p_1,p_2), \\ \nonumber
\tilde\Upsilon&=&\frac{1+(1-x)^2}{x}-\frac{2m^2}{\chi_1'}.
\end{eqnarray}
These expressions could also be inferred by using the method
of quasi-real electrons~\cite{Baier} and starting from the non-radiative
Bhabha cross section.

After integration over a hard collinear ($\vecc{k}_1\parallel\vecc{p}_1$)
photon angular phase space, the cross section of radiative Bhabha scattering
in the Born approximation is found to be
\begin{eqnarray}
\frac{\dd\sigma_0^\gamma}{\dd x\dd c}\Bigg|_{\vecc{k}_1\parallel\vecc{p}_1}
&=&\frac{4\alpha^3}{s}\left[\frac{1+(1-x)^2}{x}L_0-2\frac{1-x}{x}\right]
\\ \nonumber &\times&\left(\frac{3-3x+x^2
+2cx(2-x)+c^2(1-x(1-x))}{(1-x)(1-c)a^2}\right)^2(1+{\cal O}(\theta_0^2)),
\end{eqnarray}
where $L_0=\ln\left(\varepsilon\theta_0/m\right)^2$.
And in the case $\vecc{k}_1\parallel\vecc{p}_1'$ it reads
\begin{eqnarray}
\frac{\dd\sigma_0^\gamma}{\dd x\dd c}\Bigg|_{\vecc{k}_1\parallel\vecc{p}_1'}
&=&\frac{\alpha^3}{4s}\left[\frac{1+(1-x)^2}{x}L_0' - 2\frac{1-x}{x}\right]
\left(\frac{3+c^2}{1-c}\right)^2(1+{\cal O}(\theta_0^2)), \\ \nonumber
L_0'&=&\ln\left(\frac{\varepsilon_1'\theta_0}{m}\right)^2,\qquad
\varepsilon_1'=\varepsilon(1-x).
\end{eqnarray}

The simplest way to reproduce these results is to use the physical
gauge for the real photon which
in the beam cms sets the photon polarization vector to be a space-like
3-vector $\vecc{e}_\lambda$ having density matrix
$$\sum\limits_\lambda e^\lambda_\mu e^{\lambda *}_\nu=
\left\{
\begin{array}{cc}
0, & \mathrm{if}\,\,\mu\,\,\mathrm{or}\,\,\nu=0 \\
\delta_{\mu\nu}-n_\mu n_\nu, & \mu=\nu=1,2,3
\end{array}
\right.,
\qquad \vecc{n}=\frac{\vecc{k}_1}{\omega_1},$$
with the properties
\begin{eqnarray}
\sum\limits_\lambda |e_\lambda|^2 &=& -2,\quad
\sum\limits_\lambda |p_1e_\lambda|^2=\varepsilon^2(1-c_1^2), \\ \nonumber
\sum\limits_\lambda |p_1'e_\lambda|^2&=&\frac{t_1u_1}{s},\quad
\sum\limits_\lambda(p_1e_\lambda)(p_1'e_\lambda)^*
\stackrel{\theta \to 0}{\sim}\theta.
\end{eqnarray}
These properties enable us to omit mass terms in the calculations of
traces and, besides, to restrict ourselves to the consideration of
{\it singular} terms (see Eq.~(\ref{sing})) only, both at the Born and one--loop level.
As shown in~\cite{BCM}, this gauge is proved useful for a description
of jet production in quantum chromodynamics; it is also very well suited
to our case because it allows to simplify a lot the calculation with
respect, for instance, to the Feynman gauge. What is more, it possesses another
very attractive feature related with the structure of the correction
to be mentioned below (see Appendix).

With these tools at our disposal let us turn now to the main point.
The contributions, which survive the limit $\theta_0\to 0$, arise from
the terms containing
\begin{equation}\label{sing}
\frac{(p_1e)^2}{\chi_1^2},\quad \frac{e^2}{\chi_1},
\quad \frac{(p_1'e)^2}{\chi_1}.
\end{equation}
Other omitted terms (in particular those which do not contain
a factor $\chi_1^{-1}$) can be safely neglected since they
give a contribution of the order of $\theta_0^2$ which determines the
accuracy of our calculations
\begin{eqnarray}\label{acc}
1+{\cal O}\left(\frac{\alpha}{\pi} \theta_0^2L_s\right),
\qquad\frac{m}{\varepsilon}\ll\theta_0\ll 1.
\end{eqnarray}
In the realistic case this corresponds to an accuracy
of the order of per mille.

\section{Crossing relations}

In this and the next section we shall consider the case
$\vecc{k}_1\parallel\vecc{p}_1$.
In the case of photon emission along $p_1'$ one can get the desired
expression by using the {\it left-to-right} permutation
\begin{equation}
|M|^2_{\vecc{k}_1\parallel\vecc{p}_1'}
={\cal Q}\left(\begin{array}{c}p_1\leftrightarrow -p_1'\\
p_2\leftrightarrow -p_2'\end{array}\right)|M|^2_{\vecc{k}_1\parallel\vecc{p}_1}.
\end{equation}
%---------------------------------------------
\begin{figure}[ht!]
\unitlength=2.10pt
\special{em:linewidth 0.4pt}
\linethickness{0.4pt}
\begin{picture}(214.33,159.50)
\put(135.14,29.00){\oval(2.00,2.00)[l]}
\put(135.14,31.00){\oval(2.00,2.00)[r]}
\put(135.14,33.00){\oval(2.00,2.00)[l]}
\put(135.14,35.00){\oval(2.00,2.00)[r]}
\put(135.14,37.00){\oval(2.00,2.00)[l]}
\put(135.14,39.00){\oval(2.00,2.00)[r]}
\put(135.14,41.00){\oval(2.00,2.00)[l]}
\put(135.14,43.00){\oval(2.00,2.00)[r]}
\put(135.14,45.00){\oval(2.00,2.00)[l]}
\put(115.14,71.00){\line(4,-5){20.00}}
\put(135.14,46.00){\line(4,5){20.00}}
\put(135.14,28.00){\line(-1,-1){20.00}}
\put(135.14,28.00){\line(1,-1){20.00}}
\put(135.14,28.00){\vector(-1,-1){12.00}}
\put(155.14,8.00){\vector(-1,1){9.00}}
\put(127.14,57.51){\oval(2.00,2.33)[t]}
\put(129.14,57.67){\oval(2.00,2.67)[b]}
\put(131.14,57.51){\oval(2.00,2.33)[t]}
\put(133.14,57.67){\oval(2.00,2.67)[b]}
\put(135.14,57.51){\oval(2.00,2.33)[t]}
\put(137.14,57.67){\oval(2.00,2.67)[b]}
\put(139.14,57.51){\oval(2.00,2.33)[t]}
\put(141.14,57.67){\oval(2.00,2.67)[b]}
\put(143.14,57.51){\oval(2.00,2.33)[t]}
\put(135.81,7.00){\makebox(0,0)[cc]{$(5)$}}
\put(117.80,67.67){\vector(3,-4){1.67}}
\put(130.80,51.67){\vector(3,-4){1.67}}
\put(191.81,29.00){\oval(2.00,2.00)[l]}
\put(191.81,31.00){\oval(2.00,2.00)[r]}
\put(191.81,33.00){\oval(2.00,2.00)[l]}
\put(191.81,35.00){\oval(2.00,2.00)[r]}
\put(191.81,37.00){\oval(2.00,2.00)[l]}
\put(191.81,39.00){\oval(2.00,2.00)[r]}
\put(191.81,41.00){\oval(2.00,2.00)[l]}
\put(191.81,43.00){\oval(2.00,2.00)[r]}
\put(191.81,45.00){\oval(2.00,2.00)[l]}
\put(171.81,71.00){\line(4,-5){20.00}}
\put(191.81,46.00){\line(4,5){20.00}}
\put(191.14,28.00){\line(-1,-1){20.00}}
\put(191.14,28.00){\line(1,-1){20.00}}
\put(191.14,28.00){\vector(-1,-1){12.00}}
\put(211.14,8.00){\vector(-1,1){9.00}}
\put(179.81,62.51){\oval(2.00,2.33)[t]}
\put(181.81,62.67){\oval(2.00,2.67)[b]}
\put(183.81,62.51){\oval(2.00,2.33)[t]}
\put(185.81,62.67){\oval(2.00,2.67)[b]}
\put(187.81,62.51){\oval(2.00,2.33)[t]}
\put(189.81,62.67){\oval(2.00,2.67)[b]}
\put(191.81,62.51){\oval(2.00,2.33)[t]}
\put(193.81,62.67){\oval(2.00,2.67)[b]}
\put(195.81,62.51){\oval(2.00,2.33)[t]}
\put(197.81,62.67){\oval(2.00,2.67)[b]}
\put(199.81,62.51){\oval(2.00,2.33)[t]}
\put(201.81,62.67){\oval(2.00,2.67)[b]}
\put(203.81,62.51){\oval(2.00,2.33)[t]}
\put(186.81,53.66){\oval(2.00,3.33)[lt]}
\put(186.81,57.00){\oval(2.00,3.33)[rb]}
\put(188.81,56.66){\oval(2.00,3.33)[lt]}
\put(188.81,60.00){\oval(2.00,3.33)[rb]}
\put(190.81,59.66){\oval(2.00,3.33)[lt]}
\put(190.81,63.00){\oval(2.00,3.33)[rb]}
\put(192.81,62.66){\oval(2.00,3.33)[lt]}
\put(192.81,66.00){\oval(2.00,3.33)[rb]}
\put(194.81,65.66){\oval(2.00,3.33)[lt]}
\put(194.81,69.00){\oval(2.00,3.33)[rb]}
\put(196.81,68.66){\oval(2.00,3.33)[lt]}
\put(196.81,72.00){\oval(2.00,3.33)[rb]}
\put(191.81,7.00){\makebox(0,0)[cc]{$(6)$}}
\put(174.47,67.67){\vector(3,-4){1.67}}
\put(181.14,59.67){\vector(3,-4){1.67}}
\put(187.47,51.67){\vector(3,-4){1.67}}
\put(123.80,61.33){\oval(2.00,3.33)[lt]}
\put(123.80,64.67){\oval(2.00,3.33)[rb]}
\put(125.80,64.33){\oval(2.00,3.33)[lt]}
\put(125.80,67.67){\oval(2.00,3.33)[rb]}
\put(127.80,67.33){\oval(2.00,3.33)[lt]}
\put(127.80,70.67){\oval(2.00,3.33)[rb]}
\put(129.80,70.33){\oval(2.00,3.33)[lt]}
\put(122.80,61.67){\vector(3,-4){1.67}}
\put(149.47,63.67){\vector(3,4){1.67}}
\put(138.14,49.67){\vector(3,4){1.67}}
\put(206.14,63.67){\vector(3,4){1.67}}
\put(194.47,49.34){\vector(3,4){1.67}}
\put(26.00,29.00){\oval(2.00,2.00)[l]}
\put(26.00,31.00){\oval(2.00,2.00)[r]}
\put(26.00,33.00){\oval(2.00,2.00)[l]}
\put(26.00,35.00){\oval(2.00,2.00)[r]}
\put(26.00,37.00){\oval(2.00,2.00)[l]}
\put(26.00,39.00){\oval(2.00,2.00)[r]}
\put(26.00,41.00){\oval(2.00,2.00)[l]}
\put(26.00,43.00){\oval(2.00,2.00)[r]}
\put(26.00,45.00){\oval(2.00,2.00)[l]}
\put(6.00,71.00){\line(4,-5){20.00}}
\put(26.00,46.00){\line(4,5){20.00}}
\put(26.14,28.00){\line(-1,-1){20.00}}
\put(26.14,28.00){\line(1,-1){20.00}}
\put(26.14,28.00){\vector(-1,-1){12.00}}
\put(46.14,8.00){\vector(-1,1){9.00}}
\put(13.00,63.51){\oval(2.00,2.00)[t]}
\put(15.00,63.51){\oval(2.00,2.00)[b]}
\put(17.00,63.51){\oval(2.00,2.00)[t]}
\put(19.00,63.51){\oval(2.00,2.00)[b]}
\put(21.00,63.51){\oval(2.00,2.00)[t]}
\put(23.00,63.51){\oval(2.00,2.00)[lb]}
\put(23.00,61.51){\oval(2.00,2.00)[r]}
\put(23.00,59.51){\oval(2.00,2.00)[l]}
\put(23.00,57.51){\oval(2.00,2.00)[r]}
\put(23.00,55.51){\oval(2.00,2.00)[l]}
\put(23.00,53.51){\oval(2.00,2.00)[r]}
\put(23.00,51.51){\oval(3.00,2.00)[lt]}
\put(8.00,68.50){\vector(3,-4){1.67}}
\put(16.00,58.50){\vector(3,-4){1.67}}
\put(22.00,51.00){\vector(3,-4){1.67}}
\put(35.00,57.25){\vector(3,4){1.67}}
\put(11.80,65.33){\oval(2.00,3.33)[lt]}
\put(11.80,68.67){\oval(2.00,3.33)[rb]}
\put(13.80,68.33){\oval(2.00,3.33)[lt]}
\put(13.80,71.67){\oval(2.00,3.33)[rb]}
\put(15.80,71.33){\oval(2.00,3.33)[lt]}
\put(15.80,74.67){\oval(2.00,3.33)[rb]}
\put(26.81,7.00){\makebox(0,0)[cc]{$(3)$}}
\put(81.00,29.00){\oval(2.00,2.00)[l]}
\put(81.00,31.00){\oval(2.00,2.00)[r]}
\put(81.00,33.00){\oval(2.00,2.00)[l]}
\put(81.00,35.00){\oval(2.00,2.00)[r]}
\put(81.00,37.00){\oval(2.00,2.00)[l]}
\put(81.00,39.00){\oval(2.00,2.00)[r]}
\put(81.00,41.00){\oval(2.00,2.00)[l]}
\put(81.00,43.00){\oval(2.00,2.00)[r]}
\put(81.00,45.00){\oval(2.00,2.00)[l]}
\put(61.00,71.00){\line(4,-5){20.00}}
\put(81.00,46.00){\line(4,5){20.00}}
\put(81.14,28.00){\line(-1,-1){20.00}}
\put(81.14,28.00){\line(1,-1){20.00}}
\put(81.14,28.00){\vector(-1,-1){12.00}}
\put(101.14,8.00){\vector(-1,1){9.00}}
\put(68.00,63.51){\oval(2.00,2.00)[t]}
\put(70.00,63.51){\oval(2.00,2.00)[b]}
\put(72.00,63.51){\oval(2.00,2.00)[t]}
\put(74.00,63.51){\oval(2.00,2.00)[b]}
\put(76.00,63.51){\oval(2.00,2.00)[t]}
\put(78.00,63.51){\oval(2.00,2.00)[lb]}
\put(78.00,61.51){\oval(2.00,2.00)[r]}
\put(78.00,59.51){\oval(2.00,2.00)[l]}
\put(78.00,57.51){\oval(2.00,2.00)[r]}
\put(78.00,55.51){\oval(2.00,2.00)[l]}
\put(78.00,53.51){\oval(2.00,2.00)[r]}
\put(78.00,51.51){\oval(3.00,2.00)[lt]}
\put(63.00,68.50){\vector(3,-4){1.67}}
\put(77.80,50.33){\vector(3,-4){1.67}}
\put(90.00,57.25){\vector(3,4){1.67}}
\put(74.00,56.00){\oval(2.00,3.00)[lt]}
\put(74.00,59.00){\oval(2.00,3.00)[rb]}
\put(76.00,59.00){\oval(2.00,3.00)[lt]}
\put(76.00,62.00){\oval(2.00,3.00)[rb]}
\put(78.00,62.00){\oval(2.00,3.00)[lt]}
\put(78.00,65.00){\oval(2.00,3.00)[rb]}
\put(80.00,65.00){\oval(2.00,3.00)[lt]}
\put(80.00,68.00){\oval(2.00,3.00)[rb]}
\put(82.00,68.00){\oval(2.00,3.00)[lt]}
\put(82.00,71.00){\oval(2.00,3.00)[rb]}
\put(81.81,7.00){\makebox(0,0)[cc]{$(4)$}}
\
\
\put(17.14,105.51){\oval(2.00,2.33)[t]}
\put(19.14,105.67){\oval(2.00,2.67)[b]}
\put(21.14,105.51){\oval(2.00,2.33)[t]}
\put(23.14,105.67){\oval(2.00,2.67)[b]}
\put(25.14,105.51){\oval(2.00,2.33)[t]}
\put(27.14,105.67){\oval(2.00,2.67)[b]}
\put(29.14,105.51){\oval(2.00,2.33)[t]}
\put(31.14,105.67){\oval(2.00,2.67)[b]}
\put(33.14,105.51){\oval(2.00,2.33)[t]}
\put(80.00,116.00){\oval(2.00,2.00)[l]}
\put(80.00,118.00){\oval(2.00,2.00)[r]}
\put(80.00,120.00){\oval(2.00,2.00)[l]}
\put(80.00,128.00){\oval(2.00,2.00)[l]}
\put(80.00,130.00){\oval(2.00,2.00)[r]}
\put(80.00,132.00){\oval(2.00,2.00)[l]}
\put(60.00,158.00){\line(4,-5){20.00}}
\put(80.00,133.00){\line(4,5){20.00}}
\put(80.14,115.00){\line(-1,-1){20.00}}
\put(80.14,115.00){\line(1,-1){20.00}}
\put(80.14,115.00){\vector(-1,-1){12.00}}
\put(100.14,95.00){\vector(-1,1){9.00}}
\put(62.00,155.50){\vector(3,-4){1.67}}
\put(76.80,137.33){\vector(3,-4){1.67}}
\put(89.00,144.25){\vector(3,4){1.67}}
\put(73.00,143.00){\oval(2.00,3.00)[lt]}
\put(73.00,146.00){\oval(2.00,3.00)[rb]}
\put(75.00,146.00){\oval(2.00,3.00)[lt]}
\put(75.00,149.00){\oval(2.00,3.00)[rb]}
\put(77.00,149.00){\oval(2.00,3.00)[lt]}
\put(77.00,152.00){\oval(2.00,3.00)[rb]}
\put(79.00,152.00){\oval(2.00,3.00)[lt]}
\put(79.00,155.00){\oval(2.00,3.00)[rb]}
\put(81.00,155.00){\oval(2.00,3.00)[lt]}
\put(81.00,158.00){\oval(2.00,3.00)[rb]}
\put(25.00,116.00){\oval(2.00,2.00)[l]}
\put(25.00,118.00){\oval(2.00,2.00)[r]}
\put(25.00,120.00){\oval(2.00,2.00)[l]}
\put(25.00,122.00){\oval(2.00,2.00)[r]}
\put(25.00,124.00){\oval(2.00,2.00)[l]}
\put(25.00,126.00){\oval(2.00,2.00)[r]}
\put(25.00,128.00){\oval(2.00,2.00)[l]}
\put(25.00,130.00){\oval(2.00,2.00)[r]}
\put(25.00,132.00){\oval(2.00,2.00)[l]}
\put(5.00,158.00){\line(4,-5){20.00}}
\put(25.00,133.00){\line(4,5){20.00}}
\put(25.14,115.00){\line(-1,-1){20.00}}
\put(25.14,115.00){\line(1,-1){20.00}}
\put(25.14,115.00){\vector(-1,-1){12.00}}
\put(45.14,95.00){\vector(-1,1){9.00}}
\put(7.00,155.50){\vector(3,-4){1.67}}
\put(21.80,137.33){\vector(3,-4){1.67}}
\put(34.00,144.25){\vector(3,4){1.67}}
\put(18.00,143.00){\oval(2.00,3.00)[lt]}
\put(18.00,146.00){\oval(2.00,3.00)[rb]}
\put(20.00,146.00){\oval(2.00,3.00)[lt]}
\put(20.00,149.00){\oval(2.00,3.00)[rb]}
\put(22.00,149.00){\oval(2.00,3.00)[lt]}
\put(22.00,152.00){\oval(2.00,3.00)[rb]}
\put(24.00,152.00){\oval(2.00,3.00)[lt]}
\put(24.00,155.00){\oval(2.00,3.00)[rb]}
\put(26.00,155.00){\oval(2.00,3.00)[lt]}
\put(26.00,158.00){\oval(2.00,3.00)[rb]}
\put(22.00,159.00){\makebox(0,0)[cc]{$k_1$}}
\put(5.00,150.00){\makebox(0,0)[cc]{$p_1$}}
\put(45.00,151.00){\makebox(0,0)[cc]{$p_1'$}}
\put(30.00,121.00){\makebox(0,0)[cc]{$k$}}
\put(80.00,124.00){\circle{6.00}}
\
\
\put(203.57,148.68){\oval(2.67,2.67)[lt]}
\put(203.57,151.34){\oval(2.67,2.67)[rb]}
\put(200.90,146.01){\oval(2.67,2.67)[lt]}
\put(200.90,148.68){\oval(2.67,2.67)[rb]}
\put(198.23,143.34){\oval(2.67,2.67)[lt]}
\put(198.23,146.01){\oval(2.67,2.67)[rb]}
\put(195.57,140.68){\oval(2.67,2.67)[lt]}
\put(195.57,143.34){\oval(2.67,2.67)[rb]}
\put(192.90,138.01){\oval(2.67,2.67)[lt]}
\put(192.90,140.68){\oval(2.67,2.67)[rb]}
\put(197.22,144.23){\vector(1,1){1.11}}
\put(179.33,114.67){\oval(2.00,2.00)[r]}
\put(179.33,116.67){\oval(2.00,2.00)[l]}
\put(179.33,118.67){\oval(2.00,2.00)[r]}
\put(179.33,120.67){\oval(2.00,2.00)[l]}
\put(179.33,122.67){\oval(2.00,2.00)[r]}
\put(179.33,124.67){\oval(2.00,2.00)[l]}
\put(179.33,126.67){\oval(2.00,2.00)[r]}
\put(179.33,128.67){\oval(2.00,2.00)[l]}
\put(179.33,130.67){\oval(2.00,2.00)[r]}
\put(179.33,132.67){\oval(2.00,2.00)[l]}
\put(179.33,134.67){\oval(2.00,2.00)[r]}
\put(179.33,136.67){\oval(2.00,2.00)[l]}
\put(203.66,114.67){\oval(2.00,2.00)[r]}
\put(203.66,116.67){\oval(2.00,2.00)[l]}
\put(203.66,118.67){\oval(2.00,2.00)[r]}
\put(203.66,120.67){\oval(2.00,2.00)[l]}
\put(203.66,122.67){\oval(2.00,2.00)[r]}
\put(203.66,124.67){\oval(2.00,2.00)[l]}
\put(203.66,126.67){\oval(2.00,2.00)[r]}
\put(203.66,128.67){\oval(2.00,2.00)[l]}
\put(203.66,130.67){\oval(2.00,2.00)[r]}
\put(203.66,132.67){\oval(2.00,2.00)[l]}
\put(203.66,134.67){\oval(2.00,2.00)[r]}
\put(203.66,136.67){\oval(2.00,2.00)[l]}
\put(168.66,137.67){\line(1,0){45.67}}
\put(168.66,113.67){\line(1,0){45.67}}
\put(173.33,137.67){\vector(1,0){1.00}}
\put(184.67,137.67){\vector(1,0){1.00}}
\put(197.67,137.67){\vector(1,0){1.00}}
\put(208.33,137.67){\vector(1,0){1.00}}
\put(175.00,113.67){\vector(-1,0){1.33}}
\put(193.00,113.67){\vector(-1,0){1.33}}
\put(210.67,113.67){\vector(-1,0){1.33}}
\put(4.00,102.00){\makebox(0,0)[cc]{$-p_2$}}
\put(45.00,103.00){\makebox(0,0)[cc]{$-p_2'$}}
\put(130.57,148.68){\oval(2.67,2.67)[lt]}
\put(130.57,151.34){\oval(2.67,2.67)[rb]}
\put(127.90,146.01){\oval(2.67,2.67)[lt]}
\put(127.90,148.68){\oval(2.67,2.67)[rb]}
\put(125.23,143.34){\oval(2.67,2.67)[lt]}
\put(125.23,146.01){\oval(2.67,2.67)[rb]}
\put(122.57,140.68){\oval(2.67,2.67)[lt]}
\put(122.57,143.34){\oval(2.67,2.67)[rb]}
\put(119.90,138.01){\oval(2.67,2.67)[lt]}
\put(119.90,140.68){\oval(2.67,2.67)[rb]}
\put(124.22,144.23){\vector(1,1){1.11}}
\put(122.33,114.67){\oval(2.00,2.00)[r]}
\put(122.33,116.67){\oval(2.00,2.00)[l]}
\put(122.33,118.67){\oval(2.00,2.00)[r]}
\put(122.33,120.67){\oval(2.00,2.00)[l]}
\put(122.33,122.67){\oval(2.00,2.00)[r]}
\put(122.33,124.67){\oval(2.00,2.00)[l]}
\put(122.33,126.67){\oval(2.00,2.00)[r]}
\put(122.33,128.67){\oval(2.00,2.00)[l]}
\put(122.33,130.67){\oval(2.00,2.00)[r]}
\put(122.33,132.67){\oval(2.00,2.00)[l]}
\put(122.33,134.67){\oval(2.00,2.00)[r]}
\put(122.33,136.67){\oval(2.00,2.00)[l]}
\put(141.66,114.67){\oval(2.00,2.00)[r]}
\put(141.66,116.67){\oval(2.00,2.00)[l]}
\put(141.66,118.67){\oval(2.00,2.00)[r]}
\put(141.66,120.67){\oval(2.00,2.00)[l]}
\put(141.66,122.67){\oval(2.00,2.00)[r]}
\put(141.66,124.67){\oval(2.00,2.00)[l]}
\put(141.66,126.67){\oval(2.00,2.00)[r]}
\put(141.66,128.67){\oval(2.00,2.00)[l]}
\put(141.66,130.67){\oval(2.00,2.00)[r]}
\put(141.66,132.67){\oval(2.00,2.00)[l]}
\put(141.66,134.67){\oval(2.00,2.00)[r]}
\put(141.66,136.67){\oval(2.00,2.00)[l]}
\put(108.66,137.67){\line(1,0){45.67}}
\put(108.66,113.67){\line(1,0){45.67}}
\put(113.33,137.67){\vector(1,0){1.00}}
\put(124.67,137.67){\vector(1,0){1.00}}
\put(137.67,137.67){\vector(1,0){1.00}}
\put(148.33,137.67){\vector(1,0){1.00}}
\put(115.00,113.67){\vector(-1,0){1.33}}
\put(133.00,113.67){\vector(-1,0){1.33}}
\put(150.67,113.67){\vector(-1,0){1.33}}
\put(25.00,94.00){\makebox(0,0)[cc]{$(1)$}}
\put(81.00,95.00){\makebox(0,0)[cc]{$(2)$}}
\put(133.00,95.00){\makebox(0,0)[cc]{$(7)$}}
\put(192.00,95.00){\makebox(0,0)[cc]{$(8)$}}
\put(76.22,150.00){\vector(1,1){1.00}}
\put(21.22,150.00){\vector(1,1){1.00}}
\put(125.80,66.23){\vector(1,1){1.00}}
\put(194.00,66.80){\vector(1,1){1.00}}
\put(79.00,65.70){\vector(1,1){1.00}}
\put(13.00,69.50){\vector(1,1){1.00}}
\end{picture}
\caption{Some representatives of FD for radiative Bhabha scattering
        up to second order:{\bf\sl (1)} is the vertex insertion; {\bf\sl (2)}
        is the vacuum polarization insertion; graphs denoted by
        {\bf\sl (3),(4)} are of the L-type, {\bf\sl (5)} is of $G_1$-type,
        {\bf\sl (6)} is of $G_2$-type, {\bf\sl (7)} is of B-type and
        {\bf\sl (8)} is of P-type. }
\end{figure}
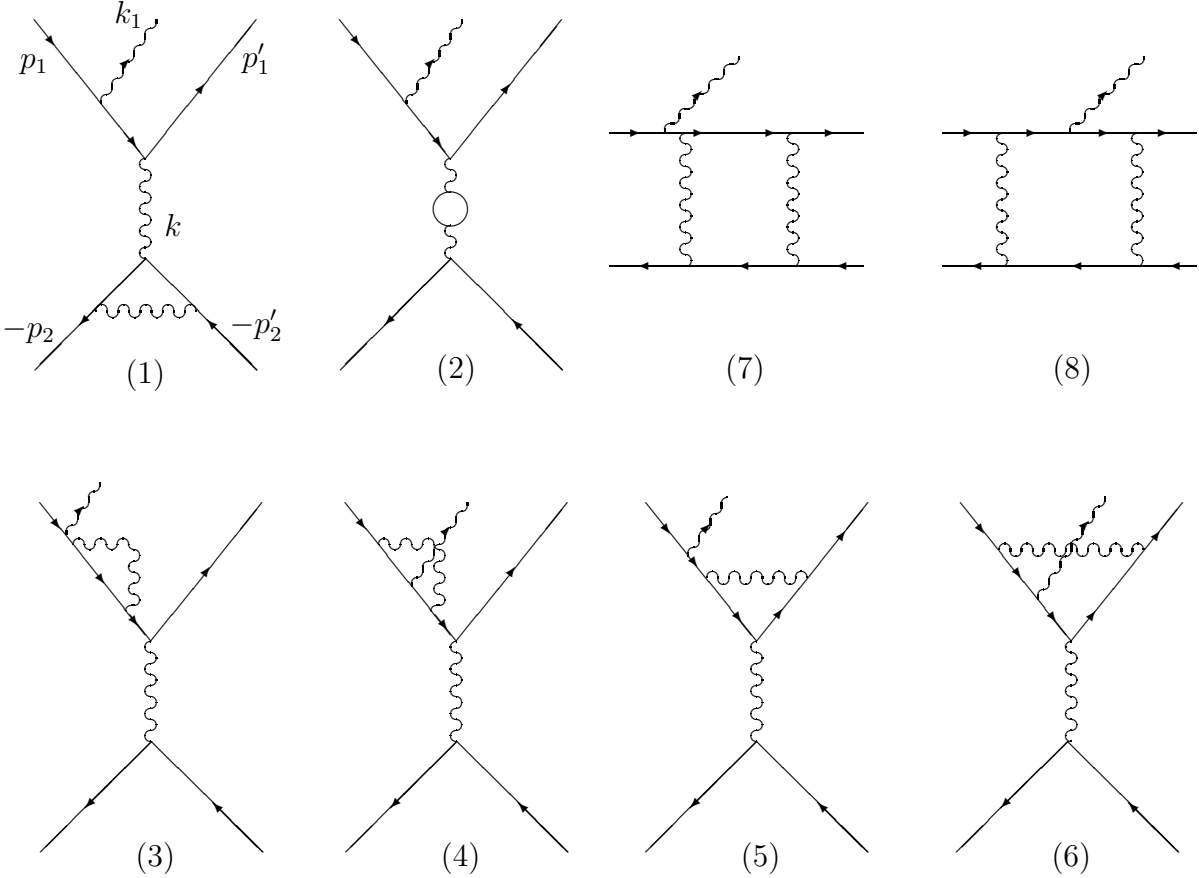

From now on we deal with scattering type amplitudes (FD) with the emission of
hard photon by initial electron.
This is possible due to the properties of the physical gauge.
The contribution of annihilation
type amplitudes may be derived by applying the momenta replacement
operation as follows:
\begin{eqnarray}
\Delta |M|^2_{\matrm{annihilation}}=\left\{{\cal Q}
(p_1'\leftrightarrow -p_2)\right\} \Delta |M|^2_{\matrm{scattering}}
\equiv \left\{Q_1\right\}
\Delta |M|^2_{\matrm{scattering}}.
\end{eqnarray}

In considering FD with two photons in the scattering channel (box FD)
one may examine only those with uncrossed photons because a contribution
of the others may be obtained by the permutation
$p_2\leftrightarrow -p_2'$. Thus the general answer becomes
\begin{equation}\label{k1p1}
|M|^2_{\vecc{k}_1\parallel \vecc{p}_1}
=\Re e\{(1+Q_1)[G+L]+\frac{1}{s_1t}(1+Q_1)(1+Q_2)[s_1t(B+P)]\},
\end{equation}
with the permutation operators acting as
$$Q_1F(s_1,t_1,s,t)=F(t,s,t_1,s_1),\quad Q_2F(s,u,s_1,u_1)=F(u,s,u_1,s_1).$$

\section{Virtual and soft photon emission in
$\vecc{k}_1\parallel\vecc{p}_1$ kinematics }

One--loop QED RC (which are described by seventy two Feynman diagrams)
can be classified out into the two gauge invariant subsets (see~Fig.1):
\begin{itemize}
\item{single photon exchange between electron and positron lines
(G,L--type);}
\item{double photon exchange between electron and positron lines
(B,P--type).}
\end{itemize}
For L-type FD (see Fig.~1(3,4)) the initial spinor $u(p_1)$ is replaced by
$(\alpha/(2\pi)) A_2 \hat{k}_1\hat{e} u(p_1)$, with
\begin{eqnarray*}
&&A_2=\frac{1}{\chi_1}\Biggl\{-\frac{\rho}{2(\rho-1)}
+ \frac{2\rho^2 - 3\rho + 2}{2(\rho - 1)^2} L_{\rho}
+\frac{1}{\rho}\left[-\Li(1-\rho) + \frac{\pi^2}{6}\right]\Biggl\}, \\
&&L_\rho=\ln\rho,\qquad \rho=\frac{\chi_1}{m^2}.
\end{eqnarray*}
The relevant contribution to the matrix element squared and
summed over spin states reads
\begin{eqnarray}
&&\Delta|M|^2_L=2^9\pi^2\alpha^4\frac{A_2}{\chi_1}\frac{s_1^3-u_1^3}{s_1 t^2}
\left[Y-\frac{2(2-x)}{1-x}W\right], \\ \nonumber
&&Y=4(p_1e)^2-\frac{x}{1-x}e^2\chi_1, \qquad W=(p_1e)^2.
\end{eqnarray}

The contribution of vertex insertion, vacuum polarization\footnote{
For realistic applications one should also add to $\Pi$ the contributions
due to $\mu$ and $\tau$ leptons and hadrons.}  and $G_1$-type
(see Fig.~1(1,2,5)) has the following form
\begin{eqnarray}
\Delta|M|^2_{\Pi,\Gamma,\Gamma_a}& =& 2^{10}\pi^2\alpha^4
\left[\Pi_t+\Gamma_t+\frac{1}{4}\Gamma_a\right]
\frac{s_1^3-u_1^3}{t^2s_1\chi_1^2} Y, \\ \nonumber
\Pi_t&=&\frac{1}{3}L_t-\frac{5}{9},\quad
\Gamma_t=(L_\lambda-1)(1-L_t)-\frac{1}{4}L_t
-\frac{1}{4}L_t^2 + \frac{\pi^2}{12},\\ \nonumber
\Gamma_a & = &-3L_t^2+4L_tL_\rho+3L_t+4L_\lambda
-2\ln(1-\rho)-\frac{\pi^2}{3} + 2\Li(1-\rho) - 4, \\ \nonumber
L_\lambda&=&\ln\frac{m}{\lambda}, \quad L_t=\ln\frac{-t}{m^2},\quad
\Li(z)=-\int\limits_0^{z}\frac{\dd x}{x}\ln(1-x).
\end{eqnarray}
Here $\lambda$ is as usual the IR cut-off parameter to be cancelled at
the end of calculus against a soft photon contribution.

For the contribution of $G_2$-type FD (see Fig.~1(6)) with four denominators
we obtain
\begin{eqnarray}
\Delta|M|^2_G&=&2^{9}\alpha^4\pi^2\frac{s_1^3-u_1^3}{t s_1\chi_1(1-x)}
\left[(J-J_1)Y + \frac{2(2-x)}{1-x} W (J_{11}-J_1+xJ_{1k}-xJ_{k})\right].
\label{16}
\end{eqnarray}
It turns out that only the scalar integral and the coefficients before
$p_1,k_1$ in the vector and tensor integrals give non-vanishing
contribution in the limit $\theta_0\to 0$
\begin{eqnarray*}
&&\int \frac{\dd^4 k}{\ii\pi^2}\frac{(1, k^\mu, k^\mu k^\nu)}{(0)(1)(2)(q)}
=(J, J_1p_1^\mu+J_kk_1^\mu, J_{11}p_1^\mu p_1^\nu+J_{kk}k_1^\mu k_1^\nu
+J_{1k}(p_1k_1)^{\mu\nu}), \\
&&(0)=k^2-\lambda^2,\ (1)=k^2-2p_1k,\ (2)=k^2-2p_1'k,
\ (q)=k^2-2k(p_1-k_1)-\chi_1, \\
&&(ab)^{\mu\nu}=a^\mu b^\nu + a^\nu b^\mu,
\end{eqnarray*}
and the terms having no $p_1$ or $k_1$ momentum in the decomposition
have been omitted for their unimportance.

The B-type FD shown in Fig.~1(7) with uncrossed legs gives
\begin{eqnarray}
\Delta|M|^2_B&=&2^9 \pi^2\alpha^4 Y\frac{1}{s_1 t \chi_1^2}
\Biggl[(u_1^3-s_1^3)s_1(B + a - b) - u_1^3s_1\left(c+a_{1'2'}+a_{1'2}
+\frac{2}{s_1}a_g\right) \nonumber \\
&+& s_1^3(c[t-u_1]+2J_0)\Biggr],
\end{eqnarray}
where the coefficients are associated with scalar, vector and tensor
integrals over loop momentum
\begin{eqnarray*}
&&\int \frac{\dd^4 k}{\ii\pi^2}\frac{(1, k^\mu, k^\mu k^\nu)}{b_0b_1b_2b_3}=
(B, B^\mu, B^{\mu\nu}), \qquad J_0=\int\frac{\dd^4k}{\ii\pi^2}
\frac{1}{b_1b_2b_3},\\
b_0&=&k^2-\lambda^2,\quad b_1=k^2+2p_1'k,\quad b_2=k^2-2p_2'k,
\quad b_3=k^2-2qk+t,\quad q=p_2'-p_2,\\
B^\mu&=&(ap_1'+bp_2'+cp_2)^\mu, \\
B^{\mu\nu}&=&a_gg^{\mu\nu}
+ a_{1'1'}p_1^{'\mu}p_1^{'\nu} + a_{22}p_2^{\mu}p_2^{\nu}
+ a_{2'2'}p_2^{'\mu}p_2^{'\nu} + a_{1'2}(p_1'p_2)^{\mu\nu}
+ a_{1'2'}(p_1'p_2')^{\mu\nu} \\
&+& a_{22'}(p_2p_2')^{\mu\nu}.
\end{eqnarray*}

For P-type FD (see Fig.~1(8)) with uncrossed photon legs we have
\begin{equation}
\Delta|M|^2_P=2^9\pi^2\alpha^4\frac{s_1^3-u_1^3}{t \chi_1(1-x)}
\left[ Y(E-E_1) + \frac{2(2-x)}{1-x} W (E_{11} - E_1 + xE_{1k} - xE_k)\right].
\end{equation}
Here we are using the definition (with tensor structures giving no 
contributions in the limit $\theta_0\to 0$ dropped)
\begin{eqnarray*}
&&\int\frac{\dd^4 k}{\ii\pi^2}\frac{(1, k^\mu, k^\mu k^\nu)}{a_0a_1a_2a_3a_4}
=(E, E_1p_1^\mu + E_kk_1^\mu, E_{11}p_1^\mu p_1^\nu + E_{kk}k_1^\mu k_1^\nu
+ E_{1k}(p_1^\mu k_1^\nu + p_1^\nu k_1^\mu)), \\
a_0&=&k^2-\lambda^2,\qquad a_1=k^2 - 2p_1k,
\qquad a_2=k^2 - 2k(p_1-k_1) - \chi_1,\\
a_3&=&k^2 + 2p_2k,\qquad a_4=k^2 - 2qk + t.
\end{eqnarray*}
Note that in the evaluating of P-type FD
we are allowed to put $k_1=xp_1$, thus keeping
only $p_1$ momentum containing terms in the decomposition.

Collecting all the contributions (for the explicit expressions
of all the coefficients see Appendix) given above
we arrive at the general expression for the virtual corrections
with $\rho=x[1+(\varepsilon\theta/m)^2]\ll s/m^2$
\begin{eqnarray} \label{main}
2\Re e\sum(M_0^*M)_{\vecc{k}_1\parallel\vecc{p}_1}
&=&\frac{2^{11}\alpha^4\pi^2}{\chi_1}F\Upsilon\Biggl\{
\frac{2-x}{1-x}\frac{w}{\Upsilon}\Phi + 2L_\lambda(2-L_t-L_{t_1}-L_s \\ \nonumber
&-&L_{s_1}+L_u+L_{u_1}) + \frac{\pi^2}{3} + \Li(x) - \frac{101}{18}
+\ln\left|\frac{\rho}{1-\rho}\right| + L_{u_1}^2 - L_t^2  \\ \nonumber
&-& L_{s_1}^2 + L_\rho\ln(1-x) + \frac{11}{3}L_t - \vartheta + \ln^2\frac{s_1}{t}
+ \frac{1}{F}\Biggl[\Pi + 3\frac{t^3-u_1^3}{s_1^2t}\ln\frac{s_1}{t} \\ \nonumber
&+&\frac{2u_1(u_1^2+s_1^2)-ts_1^2}{4t^2s_1}\ln^2\frac{u_1}{t}
+\frac{2u_1(u_1^2+t^2)-t^2s_1}{4ts_1^2}\ln^2\frac{u_1}{s_1} \\ \nonumber
&+&\frac{s_1}{2t}\ln\frac{u_1}{t} + \frac{t}{2s_1}\ln\frac{u_1}{s_1}
-\frac{3}{4}\pi^2\left(\frac{s_1}{t}+\frac{t}{s_1}\right)\Biggr]\Biggr\},
\end{eqnarray}
where we have used the following definitions
\begin{eqnarray*}
\vartheta&=&\frac{x}{\rho-x}\biggl[\Li(1-\rho)-
\frac{\pi^2}{6}+\Li(x) + L_\rho\ln(1-x)\biggr],\\
\Pi&=&\frac{s_1^3-u_1^3}{s_1t^2}\biggl[\frac{\pi}{\alpha}\left(
\frac{1}{1-\Pi_t} - 1\right) - \frac{1}{3}L_t + \frac{5}{9}\biggr]
+\frac{t^3-u_1^3}{s_1^2t}\biggl[\frac{\pi}{\alpha}\Re e\left(
\frac{1}{1-\Pi_{s_1}} - 1\right) - \frac{1}{3}L_{s_1} + \frac{5}{9}\biggr],\\
\Pi_{s_1}&=& \frac{1}{3}\left(L_{s_1}-\ii\pi\right) - \frac{5}{9},\quad
\Phi=\chi_1A_2 + t_1\chi_1(J_{11}-J_1+xJ_{1k}-xJ_{k}),\quad
w=\frac{1}{x}-\frac{1}{\rho}, \\
L_{s_1}&=&\ln\frac{s_1}{m^2},
\quad L_u=\ln\frac{-u}{m^2},\quad L_{u_1}=\ln\frac{-u_1}{m^2},
\quad L_t=\ln\frac{-t}{m^2}, \quad L_{t_1}=\ln\frac{-t_1}{m^2}.
\end{eqnarray*}

After integration over $\chi_1$ one gets additional large logs of the
form $L_0=L_s + \ln(\theta_0^2/4)$. Terms containing the last factor
have to be cancelled
against a contribution coming from the emission of hard photon
outside a narrow cone $\theta<\theta_0\ll 1$ (and supplied by the same
set of virtual and soft corrections), which was considered in~\cite{viol}.
In the case of two hard photon emission it is necessary to consider
four kinematical regions, namely when both are emitted inside/outside a cone
and one inside/another outside.

Fortunately enough, the $w$-structure, which
obviously violates factorization feature, does not contribute in~LLA
due to a cancellation of large logs in $\Phi$. What for a correction to
the above structure coming from $P$-type graph it vanishes in the sum of
FD with crossed and uncrossed photon legs (for a more comprehensive
account see Appendix).

The total expression can be obtained by summing virtual photon
emission corrections and those arising from the emission of
additional soft photon with energy exceeding no
$\Delta\varepsilon \ll \varepsilon$.

The emission of a soft photon is seen as a process factored out
of a hard subprocess (in our case the latter is exactly a hard collinear
photon emission)
so this is seemingly come into an evident conflict
with a hard collinear emission.
Nevertheless, arguments similar to those given in the paper
devoted to the problem of DIS with tagged photon~\cite{DIS} may be applied
in the present paper: the factorization of the two in the
{\it differential} cross section is present and we are, hence, allowed
to consider a soft photon emission restricted as usual by
\begin{eqnarray}\label{cnd}
\frac{\Delta\varepsilon}{\varepsilon}\ll 1.
\end{eqnarray}
Thus the soft correction can be written as
\begin{eqnarray}\label{h+s}
\sum |M|^2_{\matrm{hard+soft}}&=&\sum |M|^2_B w_{\matrm{soft}}
(\vecc{k}_1 \parallel \vecc{p}_{1}), \\ \nonumber
w_{\matrm{soft}}(\vecc{k}_1 \parallel \vecc{p}_{1})&=&-\frac{\alpha}{4\pi^2}
\int\limits_{\omega <\Delta \varepsilon}\frac{\dd^3\vecc{k}}
{\sqrt{\vecc{k}^2+\lambda^2}}
\left(-\frac{p_1}{p_1k} + \frac{p_1'}{p_1'k} + \frac{p_2}{p_2k}
- \frac{p_2'}{p_2'k}\right)^2,
\end{eqnarray}
where $M_B$ denotes the matrix element of the hard photon emission
at the Born level and in the kinematics
$\vecc{k}_1\parallel\vecc{p}_1$ it reads
\begin{eqnarray}\label{bs}
\sum|M|_B^2=\frac{2^{11}\alpha^3\pi^3}{\chi_1}\Upsilon F.
\end{eqnarray}
Now let us check the cancellation of the terms containing
$L_{\lambda}$. Indeed it takes place in the sum of contributions arising
from emission of virtual and soft real photons. To show that we bring
the soft correction into the form
\begin{eqnarray} \label{soft}
w_{\matrm{soft}}(\vecc{k}_1 \parallel \vecc{p}_{1})
&=&\frac{\alpha}{\pi}\biggl\{2\left(\ln\frac{\Delta\varepsilon}
{\varepsilon} + L_{\lambda}\right)(-2+L_s+L_{s_1}
+L_t+L_{t_1}-L_u-L_{u_1})+\frac{1}{2}(L_s^2+L_{s_1}^2 \nonumber \\
&+&L_t^2 + L_{t_1}^2-L_u^2-L_{u_1}^2) + \ln y_1(L_{u_1}-L_{s_1}
-L_{t_1})+\ln y_2(L_u-L_t-L_{s_1})  \nonumber \\
&+& \ln(y_1y_2) - \frac{2\pi^2}{3} - \frac{1}{2}\ln^2\frac{y_1}{y_2}
+\Li\left(\frac{1+c_{1'2'}}{2}\right)+\Li\left(\frac{1+c_{1'}}{2}\right)  \nonumber \\
&+&\Li\left(\frac{1-c_{2'}}{2}\right) - \Li\left(\frac{1-c_{1'}}{2}\right)
-\Li\left(\frac{1+c_{2'}}{2}\right)\biggr\},
\end{eqnarray}
where $c_i$ are the cosines of emission angles of $i$-th particle with respect
to the beam direction ($\vecc{p}_1$ in cms), $c_{1'2'}$ is the cosine
of the angle between scattered fermions in cms of the colliding particles
and $y_i$ are their energy
fractions and in the case $\vecc{k}_1\parallel\vecc{p}_1$ we have
\begin{eqnarray}
c_1'&=&c,\qquad \frac{1+c_{1'2'}}{2} = 1 - \frac{1-x}{y_1y_2},
\qquad \frac{1-c_2'}{2} = \frac{y_1(1+c)}{2y_2(1-x)}.
\end{eqnarray}

Then the cancellation of infrared singularities in the sum is evident
from comparison of Eqs.~(\ref{main},\ref{soft}).
The terms with $\ln(\Delta\varepsilon/\varepsilon)$ should be cancelled
when adding a contribution of a second hard photon having energy above
the registration threshold $\Delta\varepsilon$.

The complete expression for the correction in the case
$\vecc{k}_1\parallel\vecc{p}_1$ reads
\begin{eqnarray}\label{Res}
R&=&2\Re e\sum(M_0^*M) + |M|^2_{\matrm{soft}}=\frac{2^{11}\alpha^4\pi^2}
{\chi_1}F\Upsilon\Biggl\{\frac{2-x}{1-x}\frac{w}{\Upsilon}\Phi
+ 4\ln\left(\frac{\Delta\varepsilon}{\varepsilon}\right)
\biggl[-1 + L_{t_1} \nonumber \\
&&+\frac{1}{2}\left(-\ln(1-x) + 2\ln\frac{s}{-u}\right)\biggr]
+ \frac{11}{3}L_t
+ (L_\rho - L_t)\ln(1-x) -L_t\ln(y_1y_2) + \ln^2\frac{s_1}{-t} \nonumber \\
&& + \ln y_1\ln(1-x) + \ln(y_1y_2)\left(1+\ln\frac{-u}{s}\right)
- \frac{\pi^2}{3} + \Li(x) - \frac{101}{18} - \vartheta
+ \ln\left|\frac{\rho}{1-\rho}\right| \nonumber \\
&&-\frac{1}{2}\ln^2\frac{y_1}{y_2} + \ln(1-x)\ln\frac{-u}{s}
+ \Li\left(\frac{1+c_{1'2'}}{2}\right)
+ \Li\left(\frac{1+c_{1'}}{2}\right)
+ \Li\left(\frac{1-c_{2'}}{2}\right) \nonumber \\
&& - \Li\left(\frac{1-c_{1'}}{2}\right)
- \Li\left(\frac{1+c_{2'}}{2}\right)
+ \frac{1}{F}\Biggl[\Pi + 3\frac{t^3-u_1^3}{s_1^2t}\ln\frac{s_1}{-t}
+\frac{2u_1(u_1^2+s_1^2)-ts_1^2}{4t^2s_1}\ln^2\frac{u_1}{t} \nonumber \\
&&+\frac{2u_1(u_1^2+t^2)-t^2s_1}{4ts_1^2}\ln^2\frac{-u}{s}
+\frac{s_1}{2t}\ln\frac{u_1}{t} + \frac{t}{2s_1}\ln\frac{-u}{s}
-\frac{3}{4}\pi^2\left(\frac{s_1}{t}+\frac{t}{s_1}\right)\Biggr]\Biggr\}, \\ \nonumber
&&\dd\sigma(\vecc{k}_1\parallel\vecc{p}_1)=\frac{1}{2^{11}\pi^5s}R\dd\Gamma.
\end{eqnarray}

\section{ Kinematics $\vecc{k}_1\parallel\vecc{p}_1'$ }

We put here a set of replacements one can use in order to obtain
the modulus of matrix element squared and summed over spin states
for the case $ \vecc{k}_1\parallel \vecc{p}_1'$, starting from the analogous
expression for $\vecc{k}_1\parallel \vecc{p}_1$ (Eq.~(\ref{k1p1})) and using
the replacement of momenta
$p_1\leftrightarrow -p_1', p_2\leftrightarrow -p_2'$. The last operation
results in the following substitutions:
\begin{eqnarray} \label{trans}
x &\to& -\frac{x}{1-x}, \nonumber\\
\chi_1 &\to& -\chi_{1}',\nonumber\\
s &\leftrightarrow& s_1, \\ \nonumber
u &\leftrightarrow& u_1, \\ \nonumber
t \to t &,&  t_1 \to t_1.
\end{eqnarray}
Then under these permutations the expression for virtual
corrections given in Eq.~(\ref{main}) gets transformed yielding the
following result for the
collinear domain $\vecc{k}_1\parallel\vecc{p}_1'$
\begin{eqnarray}
2\Re e\sum(M_0^*M)_{\vecc{k}_1\parallel\vecc{p}_1'}&=&\frac{2^{11}\alpha^4\pi^2}
{\chi_1'}\tilde F\tilde\Upsilon \Biggl\{
\frac{2-x}{1-x}\frac{\tilde w}{\tilde\Upsilon}\tilde\Phi
+ 2L_\lambda(2-L_t-L_{t_1}-L_s
 \nonumber\\
&-&L_{s_1}+L_u+L_{u_1}) + \frac{\pi^2}{3}
+ \Li\left(\frac{-x}{1-x}\right)-\frac{101}{18}
+\ln\left(\frac{\xi}{\xi+1}\right) + L_{u}^2   \nonumber\\
&-& L_t^2 - L_{s}^2 - L_\xi\ln(1-x) + \frac{11}{3}L_t + \ln^2\frac{s}{-t}
+ \frac{1}{\tilde F}\Biggl[\Pi + 3\frac{t^3-u^3}{s^2t}\ln\frac{s}{-t}
 \nonumber \\
&+&\frac{2u(u^2+s^2)-ts^2}{4t^2s}\ln^2\frac{u}{t}
+\frac{2u(u^2+t^2)-t^2s}{4ts^2}\ln^2\frac{-u}{s}
+ \frac{s}{2t}\ln\frac{u}{t} - \tilde\vartheta\nonumber \\
&+&\frac{t}{2s}\ln\frac{-u}{s}
-\frac{3}{4}\pi^2\left(\frac{s}{t}+\frac{t}{s}\right)\Biggr]\Biggr\},
\end{eqnarray}
with
\begin{eqnarray*}
\tilde\Pi&=&\frac{s^3-u^3}{st^2}\biggl[\frac{\pi}{\alpha}\left(
\frac{1}{1-\Pi_t} - 1\right) - \frac{1}{3}L_t + \frac{5}{9}\biggr]
+\frac{t^3-u^3}{s^2t}\biggl[\frac{\pi}{\alpha}\Re e\left(
\frac{1}{1-\Pi_{s}} - 1\right) - \frac{1}{3}L_{s} + \frac{5}{9}\biggr],\\
\tilde F&=&\left(\frac{s}{t}+\frac{t}{s}+1\right)^2, \qquad
\tilde w = -\frac{1-x}{x}+\frac{1}{\xi}, \qquad \xi=\frac{\chi_1'}{m^2}
\end{eqnarray*}
and $\tilde \Phi,\tilde\vartheta$ derived upon applying a set of replacements
from Eq.~(\ref{trans}) on the quantities $\Phi,\vartheta$.

The contribution from the soft photon emission is described by
\begin{eqnarray}
w_{\matrm{soft}}(\vecc{k}_1 \parallel \vecc{p}_{1}') &=&  \frac{\alpha}{\pi} \,
\biggl[4(\ln\frac{\Delta\varepsilon}{\varepsilon}
+ L_\lambda)\left(-1 + L_s + \ln\frac{1-c}{1+c} + \frac{1}{2}\ln(1-x)\right)
+ L_s^2 + 2L_s \ln\frac{1-c}{1+c} \nonumber\\
&-& \frac{1}{2}\ln^2(1-x) + \ln(1-x) + \ln^2\frac{1-c}{2}
- \ln^2\frac{1+c}{2} - \frac{2\pi^2}{3} \\ \nonumber
&+& 2\Li\left(\frac{1+c}{2}\right) - 2 \Li\left(\frac{1-c}{2}\right)\biggr].
\end{eqnarray}

The total correction for the case $\vecc{k}_1\parallel\vecc{p}_1'$
has the following form
\begin{eqnarray}
\tilde R&=&2\Re e\sum(M_0^*M) + |M|^2_{\matrm{soft}}=\frac{2^{11}\alpha^4\pi^2}
{\chi_1'}\tilde F\tilde\Upsilon\Biggl\{\frac{2-x}{1-x}\frac{\tilde w}
{\tilde\Upsilon}\tilde\Phi + 4\ln\left(\frac{\Delta\varepsilon}{\varepsilon}
\right)\biggl(- 1 + L_{s}  \nonumber \\
&+& \frac{1}{2}\ln(1-x) + \ln\frac{1-c}{1+c}\biggr)
+ \frac{\pi^2}{3} + \Li\left(\frac{-x}{1-x}\right)
-\frac{101}{18}+\ln\left(\frac{\xi}{\xi+1}\right)
-2\ln^2(1-x)  \nonumber \\
&+& \frac{11}{3}L_{t} - L_\xi\ln(1-x) + \ln^2\frac{s}{-t}
- \frac{2}{3}\pi^2 +\ln(1-x) - \tilde\vartheta
+ 2\Li\left(\frac{1+c}{2}\right) \nonumber \\
&-&2\Li\left(\frac{1-c}{2}\right)
+ \frac{1}{\tilde F}\biggl[\tilde\Pi + 3\frac{t^3-u^3}{s^2t}\ln\frac{s}{-t}
+ \frac{1}{4t^2s}\ln^2\left(\frac{u}{t}\right)
(2u(u^2+s^2)-ts^2)  \nonumber \\
&+& \frac{1}{4ts^2}\ln^2\left(\frac{-u}{s}\right)(2u(u^2+t^2)-t^2s)
+ \frac{s}{2t}\ln\frac{u}{t} + \frac{t}{2s}\ln\frac{-u}{s}
- \frac{3}{4}\pi^2\left(\frac{s}{t}+\frac{t}{s}\right)
\biggr]\Biggr\}, \\ \nonumber
&&\dd\sigma(\vecc{k}_1\parallel\vecc{p}_1')=\frac{1}{2^{11}\pi^5s}
\tilde R\dd\Gamma.
\end{eqnarray}

Performing the integration over a hard photon angular phase space (inside
a narrow cones) we put the RC to the cross section coming from
virtual and soft real additional photons valid to a logarithmic accuracy
in the form
\begin{eqnarray}\label{rat}
\frac{\dd\sigma^{\gamma(V+S)}}{\dd x\dd c}=
\frac{\dd\sigma_0^\gamma}{\dd x\dd c}\frac{\alpha}{\pi}\left[
C\frac{\Delta\varepsilon}{\varepsilon} + L_t\Xi_L + \Xi\right].
\end{eqnarray}
In the Fig.~2(a,b) given are the ratio of $\Xi/(L_t\Xi_L)$ versus $x$ for
the collinear kinematics considered above.

\section{Two hard photon emission and results in LLA}

Turning to the structure of the result obtained,
it should be noted that all the terms quadratic in {\it large} logarithms
$L_{t_1}\sim L_{s_1}\sim L_u\gg L_\rho$ are mutually cancelled out
as it should be.

From the formula~(\ref{Res}) it immediately follows that
(upon doing an integration over a hard photon angular
(within a narrow cone) phase space)
the $w$-term that is not proportional to $\Upsilon$,
which is in fact the kernel of the non-singlet electron structure function,
is not dangerous in the sense of a feasible violation of the
expected Drell--Yan form of the cross section, because it does contribute
only at next-to-leading order.

Performing the above mentioned integration and confining
ourselves to LLA we get for the sum of virtual and soft photons %(see Eq.~\ref{rat})
\begin{eqnarray}\label{4}
\frac{\dd\sigma^{\gamma(S+V)}}{\dd x\dd c}=
\frac{\dd\sigma^\gamma_0}{\dd x\dd c}\frac{\alpha}{\pi}L\left[
4\ln\frac{\Delta\varepsilon}{\varepsilon} + \frac{11}{3} - \frac{1}{2}\ln(1-x)
 - \ln(y_1y_2)\right].
\end{eqnarray}

The LLA contribution coming from the emission of second hard photon
with total energy exceeding $\Delta\varepsilon$ consists of a part
corresponding to the case in which both hard photons (with total
energy $\varepsilon x$) are emitted by initial electron~\cite{2h}
\begin{eqnarray}\label{2ha}
\frac{\dd\sigma^{2\gamma}}{\dd x\dd c}
&=&\frac{\dd\sigma_0^\gamma}{\dd x\dd c}\frac{\alpha}{\pi}
L\biggl[\frac{x{\cal P}^{(2)}_\Theta(1-x)}{4(1+(1-x)^2)}
 + \frac{1}{2}\ln(1-x)
- \ln\frac{\Delta\varepsilon}{\varepsilon} - \frac{3}{4}\biggr], \\ \nonumber
P^{(2)}_{\Theta}(z) &=& 2\biggl[ \frac{1+z^2}{1-z}\biggl(
2\ln(1-z)-\ln z + \frac{3}{2}\biggr) + \frac{1+z}{2}\ln z - 1 + z \biggr],
\end{eqnarray}
and the remaining part which describes the emission of second hard
photon along scattered electron and positrons. The latter,
upon combining with the part of contributions of soft and virtual photons
to our process
\begin{eqnarray*}
\frac{\dd\sigma^\gamma_0}{\dd x\dd c}\frac{3\alpha}{\pi}L\left[
\ln\frac{\Delta\varepsilon}{\varepsilon} + \frac{3}{4}\right]\, ,
\end{eqnarray*}
may be represented via electron structure function in the spirit
of the Drell--Yan approach
\begin{eqnarray}\label{master}
\langle\frac{\dd \sigma_0^\gamma}{\dd x\dd c}\rangle
\Bigg|_{\vecc{k}_1\parallel\vecc{p}_1}
&=&\frac{\alpha}{2\pi}\frac{1+(1-x)^2}{x}L_0\int\dd z_2\dd z_3\dd z_4
{\cal{D}}(z_2){\cal{D}}(z_3){\cal{D}}(z_4)  \label{a1}\\ \nonumber
&\times&\frac{\dd\sigma_0(p_1(1-x),z_2p_2;q_1,q_2)}{\dd c},
\end{eqnarray}
with the non-singlet structure function ${\cal{D}}(z)$~\cite{KF}
\begin{eqnarray}
{\cal{D}}(z)&=&\delta(1-z)+\frac{\alpha}{2\pi}L{\cal{P}}^{(1)}(z)
+\biggl(\frac{\alpha}{2\pi}L\biggr)^2\frac{1}{2!}\,
{\cal{P}}^{(2)}(z)+ \ldots\,, \\ \nonumber
P^{(1,2)}(z) &=& \lim_{\Delta\to 0}\biggl\{\delta(1-z)P^{(1,2)}_{\Delta}
+ \Theta(1-\Delta-z)P^{(1,2)}_{\Theta}(z)\biggr\},  \\ \nonumber
P^{(1)}_{\Delta} &=& 2\ln\Delta + \frac{3}{2}\, , \quad
P^{(1)}_{\Theta}(z) = \frac{1+z^2}{1-z}\, ,\quad
P^{(2)}_{\Delta} = \biggl(2\ln\Delta + \frac{3}{2}\biggr)^2
- \frac{2\pi^2}{3}\, , \ \ldots
\end{eqnarray}

%aggiunta
These functions describe the emission of (real and virtual) photons both by 
final electron and by positrons. The multiplier before the integral 
stands for the emission of an hard photon by the initial electron.
Thus Eq.~(\ref{master}) actually represents the partially integrated
Drell--Yan form of the cross section. Quite the same arguments are 
applicable to the second case in which an hard photon is emitted by the final 
electron.

The cross section of the hard sub-process $e(p_1z_1)+\bar e(p_2z_2)\to
e(q_1)+\bar e(q_2)$ entering Eq.~(\ref{master}) has the form
\begin{eqnarray}
\frac{\dd\sigma_0(z_1p_1,z_2p_2;q_1,q_2)}{\dd c}
=\frac{8\pi\alpha^2}{s}\left[\frac{z_1^2+z_2^2
+z_1z_2+2c(z_2^2-z_1^2)+c^2(z_1^2+z_2^2-z_1z_2)}{z_1(1-c)
(z_1+z_2+c(z_2-z_1))^2}\right]^2.
\end{eqnarray}
The momenta of scattered electron $q_1$ and positron $q_2$ are completely
determined by the energy-momentum conservation law
\begin{eqnarray*}
&&q_1^0=\varepsilon\frac{2z_1z_2}{z_1+z_2+c(z_2-z_1)},\qquad q_1^0+q_2^0=\varepsilon
(z_1+z_2),\\
&&c=\cos\widehat{\vecc{q}_1,\vecc{p}}_1,
\qquad z_1\sin\widehat{\vecc{q}_1,\vecc{p}}_1=
z_2\sin\widehat{\vecc{q}_2,\vecc{p}}_1.
\end{eqnarray*}
In general their energies differ from those detected in experiment
$\varepsilon_1',\varepsilon_2'$, namely $$\varepsilon_1'=q_1^0z_3,\qquad
\varepsilon_2'=q_2^0z_4,$$ whereas the emission angles are the same in LLA.

Collecting the two expressions presented in Eqs.~(\ref{4},\ref{2ha})
one can rewrite the result in LLA as
\begin{eqnarray}\label{36}
\frac{\dd\sigma^\gamma}{\dd x\dd c}\Bigg|_{\vecc{k}_1\parallel\vecc{p}_1}
&=&\left(\frac{\dd\sigma^\gamma_0}{\dd x\dd c}\right)_{\vecc{k}_1
\parallel\vecc{p}_1}\left\{1+\delta_1\right\},  \nonumber \\
\delta_1=\left(\displaystyle{
\frac{\langle\frac{\dd \sigma_0^\gamma}{\dd x\dd c}\rangle}
{\frac{\dd\sigma^\gamma_0}{\dd x\dd c}}}\right)_{\vecc{k}_1\parallel\vecc{p}_1}
- 1 &+& \frac{\alpha}{\pi}L\biggl[\frac{2}{3}-\ln(y_1y_2)
+\frac{x{\cal P}^{(2)}_\Theta(1-x)}{4(1+(1-x)^2)}\biggr].
\end{eqnarray}
For the case $\vecc{k}_1\parallel\vecc{p}_1'$ the correction is found to be
\begin{eqnarray}\label{37}
\frac{\dd\sigma^\gamma}{\dd x\dd c}\Bigg|_{\vecc{k}_1\parallel\vecc{p}_1'}
&=&\left(\frac{\dd\sigma^\gamma_0}{\dd x\dd c}\right)_{\vecc{k}_1
\parallel\vecc{p}_1'}\left\{1+\delta_{1'}\right\},  \nonumber \\
\delta_{1'}&=&\left(\displaystyle{
\frac{\langle\frac{\dd \sigma_0^\gamma}{\dd x\dd c}\rangle}
{\frac{\dd\sigma^\gamma_0}{\dd x\dd c}}}\right)_{\vecc{k}_1\parallel\vecc{p}_1'}
-1 + \frac{\alpha}{\pi}L\left[\frac{2}{3}
+\frac{x{\cal P}^{(2)}_\Theta(1-x)
}{4(1+(1-x)^2)}\right], \nonumber \\
\langle\frac{\dd \sigma_0^\gamma}{\dd x\dd c}\rangle
\Bigg|_{\vecc{k}_1\parallel\vecc{p}_1'}
&=&\frac{\alpha}{2\pi}\frac{1+(1-x)^2}{x}L_0'\int\dd z_1\dd z_2\dd z_4
{\cal{D}}(z_1){\cal{D}}(z_2){\cal{D}}(z_4) \label{a2} \\ \nonumber
&\times&\frac{\dd\sigma_0(z_1p_1,z_2p_2;q_1,q_2)}{\dd c},
\end{eqnarray}
with $ L_0'=L_0+2\ln(1-x)$.

For the case when the energies of scattered fermions are not detected
the expressions~(\ref{a1},\ref{a2}) may be simplified due to
$\int\dd z{\cal D}(z)=1$ and $z_3,z_4$--independence of the integrand in
$\vecc{k}_1\parallel\vecc{p}_1$ kinematics ($z_4$--independence in
$\vecc{k}_1\parallel\vecc{p}_1'$ case).

The $x$--dependence of $\delta_1$ are shown in the Fig.~3
for different values of the cosine of scattering angle $c$.
For a hard photon emission by final particles
the correction $\delta_1'$ strongly depends on the experimental conditions
of particles detection: the energy thresholds of detection
of scattered fermions. This dependence for $\delta_1$ is much more weaker,
namely about $1\%$.

In conclusion let us recapitulate the results given in Eqs.~(\ref{36},\ref{37}). 
They both respect the
Drell--Yan form for a cross section in~LLA. Nevertheless a certain
deviation away from RG structure function representation at a second order
of PT in $\vecc{k}_1\parallel\vecc{p}_1$ kinematics is observed.
The term destroying expectations based on RG approach
comes from definite contribution of a soft photon emission,
the term with $\ln(y_1y_2)$ in Eq.~(\ref{36}) which cannot be
included into the structure function approach.
Its appearance is presumably a mere consequence of a complicate
kinematics of $2\to 3$ type hard subprocess (see~\cite{viol});
        for such a kind of processes
the validity of the Drell--Yan form for a cross section was
not proved so far.
Another possible way out is a careful analysis of a {\it conflict} between
a soft and hard collinear photon emission. We have used the factorized
form of a soft photon emission~(\ref{h+s}) under the condition~(\ref{cnd}).
But, to the moment, this representation in the peculiar case at hand
is not rigorously proved as well.

The accuracy of our calculations of virtual and soft photon corrections
is determined by the omitted terms of the order of
\begin{eqnarray}
1+{\cal O}\left(\theta_0^2\frac{\alpha}{\pi}L_s,\frac{m^2}{s}
\frac{\alpha}{\pi}L_s\right),
\end{eqnarray}
which corresponds to a per mille level.
The accuracy of the correction coming from two hard photon emission
is determined by ${\cal O}((\alpha/\pi)\ln(4/\theta_0^2))$ and at $1\%$ level.

\subsection*{Acknowledgments}

We thank A.B.~Arbuzov for a critical reading of the manuscript,
many valuable comments and participating at the early stage of the
investigation. We are also indebted to L.~Lipatov for many discussions
elucidating factorization issues.
One of us (EAK) is grateful to the physical department of Insubria
University (Como) for a warm hospitality during accomplishment of
the final part of this work and to the Landau Network-Centro Volta
grant for financial support.
The support of EAK by INTAS grant 93-1867 ext. and of EAK and BGS
by Russian Foundation for Basic Research grant 99-02-17730 is 
acknowledged.

\section*{Appendix}

\setcounter{equation}{0}
\renewcommand{\theequation}{A.\arabic{equation}}

Here we give the expressions for the quantities associated with
G-type integrals:
\begin{eqnarray}
J &=& -\frac{1}{\chi_1 t_1} \left[ - 2 L_{\lambda} L_{t_1}
+ 2 L_{t_1}L_{\rho} - L_t^2 - 2\Li(x) - \frac{\pi^2}{6} \right],\nonumber\\
J_1&=&\frac{1}{t_1\chi_1}\int\limits_{0}^{\rho}
\frac{\dd z}{1-z}\frac{\ln z}{1-\lambda z}=\frac{A}{t_1\chi_1}
\left(1+\frac{x}{\rho-x}\right)=\frac{A+\vartheta}{t_1\chi_1}, \nonumber \\
J_k&=&-\frac{1}{t_1\chi_1\rho}\int\limits_{0}^{\rho}
\frac{\dd z}{1-z}
\frac{z\ln z}{1-\lambda z}, \nonumber \\
J_{11}&=& - \frac{1}{t_1\chi_1}\int\limits_0^{\rho} \frac{\dd z}
{(1-z)(1-\lambda z)}\left(1 + \frac{z \ln z}{1-z} \right), \\ \nonumber
J_{1k}&=& \frac{1}{t_1\chi_1\rho}\int\limits_0^{\rho} \frac{z\dd z}
{(1-z)(1-\lambda z)}\left(1 + \frac{z \ln z}{1-z} \right), \\ \nonumber
A&=&\Li(1-\rho)-\frac{\pi^2}{6}+\Li(x)
+L_\rho\ln(1-x),\quad \lambda=\frac{x}{\rho},\quad \rho=\frac{\chi_1}{m^2}.
\end{eqnarray}
In the limit $\rho\gg 1$ we have
\begin{eqnarray*}
\Phi=\chi_1A_2+t_1\chi_1(J_{11}-J_1+xJ_{1k}-xJ_{k})=-\frac{1}{2}
+{\cal O}(\rho^{-1})
\end{eqnarray*}
and that is the reason why $w$-structure does contribute only to
next-to-leading terms.

In general the expression for 5-denominator one--loop scalar, vector
and tensor integrals are some complicate functions of five independent
kinematical invariants (in the derivation we extensively
use the technique developed in~\cite{V}). In the limit $m^2\ll \chi_1
\ll s\sim -t$ they may be considerably simplified because of
singular $1/\chi_1$ terms only kept:
\begin{eqnarray}
E & = & \frac{1}{s_1}D_{0124}+\frac{1}{t}D_{0123},\nonumber \\
E_1 &=& -xE_k = \frac{1}{2\chi_1}\left(D_{0134} - (1-x)D_{0234} - xD_{1234}
+ \chi_1 E\right), \nonumber \\
D_{0124} & = & \frac{1}{x t_1 \chi_1}\left[L_{\rho}^2
+ 2 L_{\rho}\ln\frac{x}{1-x} - \ln^2\frac{x}{1-x} - \frac{2\pi^2}{3}\right],
\\ \nonumber
\Re e D_{0123} & = & \frac{1}{s\chi_1}\left[L_{s_1}^2 - 2L_{s_1} L_{\rho}
- 2L_s L_{\lambda} + \frac{\pi^2}{6}+2\Li(x)\right], \\ \nonumber
\Re e D_{0234} & = & \frac{1}{s_1 t}\left[L_{s_1}^2
+ 2 L_{s_1}L_\lambda - 2L_\rho L_{s_1} + 2L_{s_1}L_t
- \frac{5\pi^2}{6}\right], \\ \nonumber
\Re e D_{0134} & = & \frac{1}{s t}\left[L_{s}^2
+ 2 L_{s}L_\lambda - 2(L_{t_1}+\ln(x)) L_{s} + 2L_{s}L_t
+ \frac{7\pi^2}{6}\right], \\ \nonumber
\Re e D_{1234} & = & -\frac{1}{s_1xt_1}\left[-L_{s}^2
+ 2 L_{s}(L_{t_1} + \ln(x)) + 2L_{s_1}L_{\lambda} - \frac{7\pi^2}{6}\right] . \\ \nonumber
\end{eqnarray}
The structure $E_{11}+xE_{1k}$ has the form $1/(s\chi_1)f(x,\chi_1)$ and
will vanish after performing the operation $(1+Q_2)s_1tP$
given in ~(\ref{k1p1}) which yields a contribution of $P$-type
graphs with crossed and uncrossed photon legs.

The following coefficient for the scalar
integral is obtained in the calculation of B-type FD:
\begin{equation}
B = \frac{1}{s_1 t}\left[L_{s_1}^2+2L_{s_1} L_\lambda
-2L_{s_1}L_\rho + 2L_{s_1}L_t + \frac{\pi^2}{6}\right].
\end{equation}
For the vector integral coefficients we get
\begin{eqnarray}
a & = & -\frac{1}{2s_1u_1 t}\left[-\pi^2 s_1 + 2u_1\Li(1-\rho)-s_1 L_t^2+
t L_{s_1}^2-2t L_{s_1} L_t\right], \nonumber\\
b & = & -\frac{1}{2s_1 t}\left[\frac{2\pi^2}{3} + 2\Li(1-\rho)-2 L_{s_1}^2+
4L_{s_1}L_{\rho}-2L_{s_1} L_t\right], \\ \nonumber
c & = & \frac{1}{2s_1 u_1 t}\left[2u_1 \Li(1-\rho) + \frac{\pi^2}{6}(4u_1+6t)
+ (t-2u_1)L_{s_1}^2 - s_1 L_t^2 + 4u_1 L_{s_1}L_{\rho}
+ 2s_1 L_{s_1} L_t\right].
\end{eqnarray}
The relevant quantities for tensor B-type integrals are:
\begin{eqnarray}
a_{1'2'} & = & \frac{1}{s_1 t}
\left(\frac{\rho}{\rho-1}L_{\rho} - L_t\right), \qquad
a_g=-\frac{1}{4u_1}[(L_{s_1}-L_t)^2 + \pi^2],    \nonumber \\
a_{1'2} & = & -\frac{1}{2u_1^2}[(L_t-L_{s_1})^2 + \pi^2] +
\frac{1}{t u_1}(L_{s_1}-L_t)-\frac{1}{s_1 t}
\left(\frac{\rho}{\rho-1}L_{\rho} - L_{s_1}\right), \nonumber \\
J_0&=&\frac{1}{s_1}\left[\frac{3}{2}L_{s_1}^2-2L_{s_1}L_{\rho}
- \Li(1-\rho)-\frac{4\pi^2}{3}\right].
\end{eqnarray}

As has been mentioned in the text, the physical gauge exploited
provides a direct extraction of the kernel of the structure function
out of the traces both in the tree- and loop-level amplitudes.
The pattern emerging
\begin{eqnarray}
(\hat p_1 - \hat k_1+m)\hat e(\hat p_1+m)\hat e
(\hat p_1 - \hat k_1+m)&=& 4(p_1e)^2 (\hat p_1 - \hat k_1)
- e^2\chi_1\hat k_1     \nonumber \\
 &\approx& (1-x)Y\hat p_1, \\ \nonumber
\hat k_1\hat e(\hat p_1+m)\hat e(\hat p_1-\hat k_1+m)
&\approx& (1-x)\left(2\frac{2-x}{1-x}W - Y\right)\hat p_1
\end{eqnarray}
shows this clearly.

\newpage

\section*{Figure captions}

\vspace{.5cm}

\noindent
{\bf Figure~2.}\\
The ratio $\displaystyle{\frac{\Xi}{L_t\Xi_L}}$ (see Eq.~(\ref{rat}))
versus $x=\displaystyle{\frac{\omega_1}{\varepsilon}}$ for the case: \\
a) $\vecc{k}_1\parallel\vecc{p}_1$. \\
b) $\vecc{k}_1\parallel\vecc{p}_1'$.
\vspace{1cm}

\noindent
{\bf Figure~3.} \\
The $x$--dependence of $\delta_1$ (see Eq.~(\ref{36})) \\
%b) $\delta_{1'}$ (see Eq.~(\ref{37}))\\
for different values of the cosine of scattering angle $c$.

\vspace{1cm}\noindent
Other parameters chosen are: $\theta_0=0.1,\ \varepsilon=1$ GeV.

\begin{minipage}{12cm}
\begin{center}
\mbox{\psfig{file=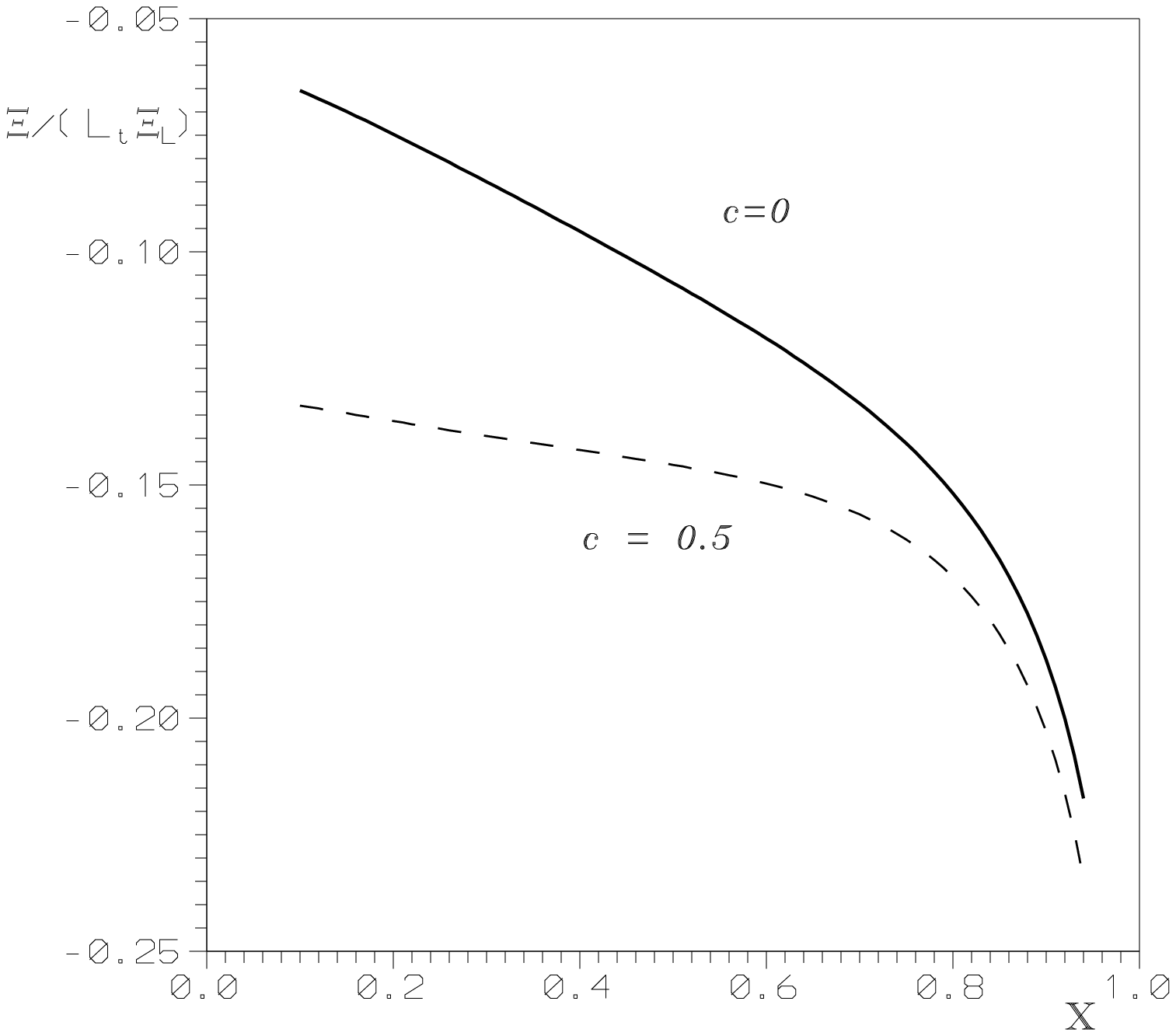,height=10cm,angle=0}}
\end{center}
\end{minipage}
\nopagebreak

\begin{minipage}{12cm}
\begin{center}
\mbox{\psfig{file=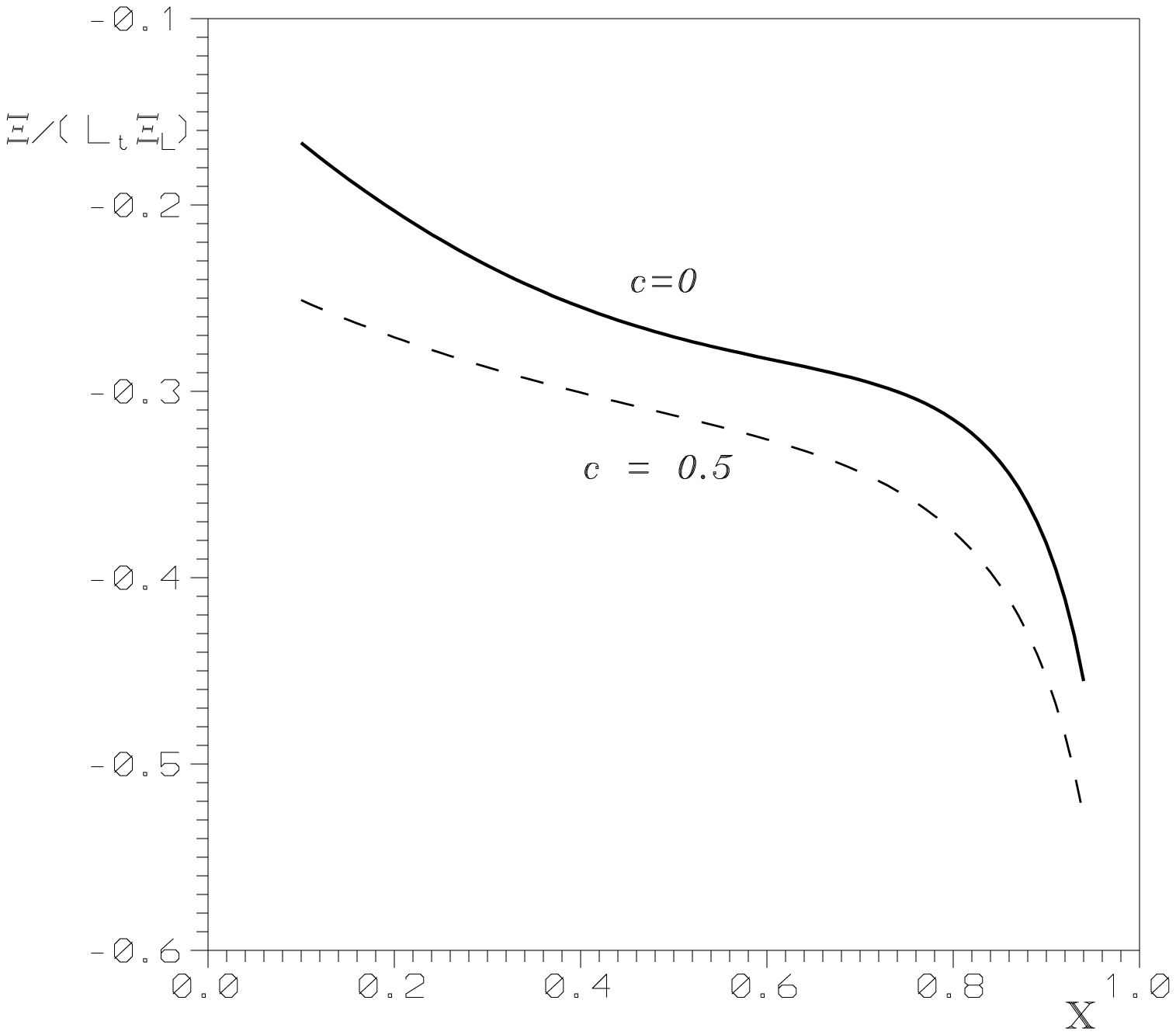,height=10cm,angle=0}}
\end{center}
\end{minipage}

\begin{minipage}{12cm}
\begin{center}
\mbox{\psfig{file=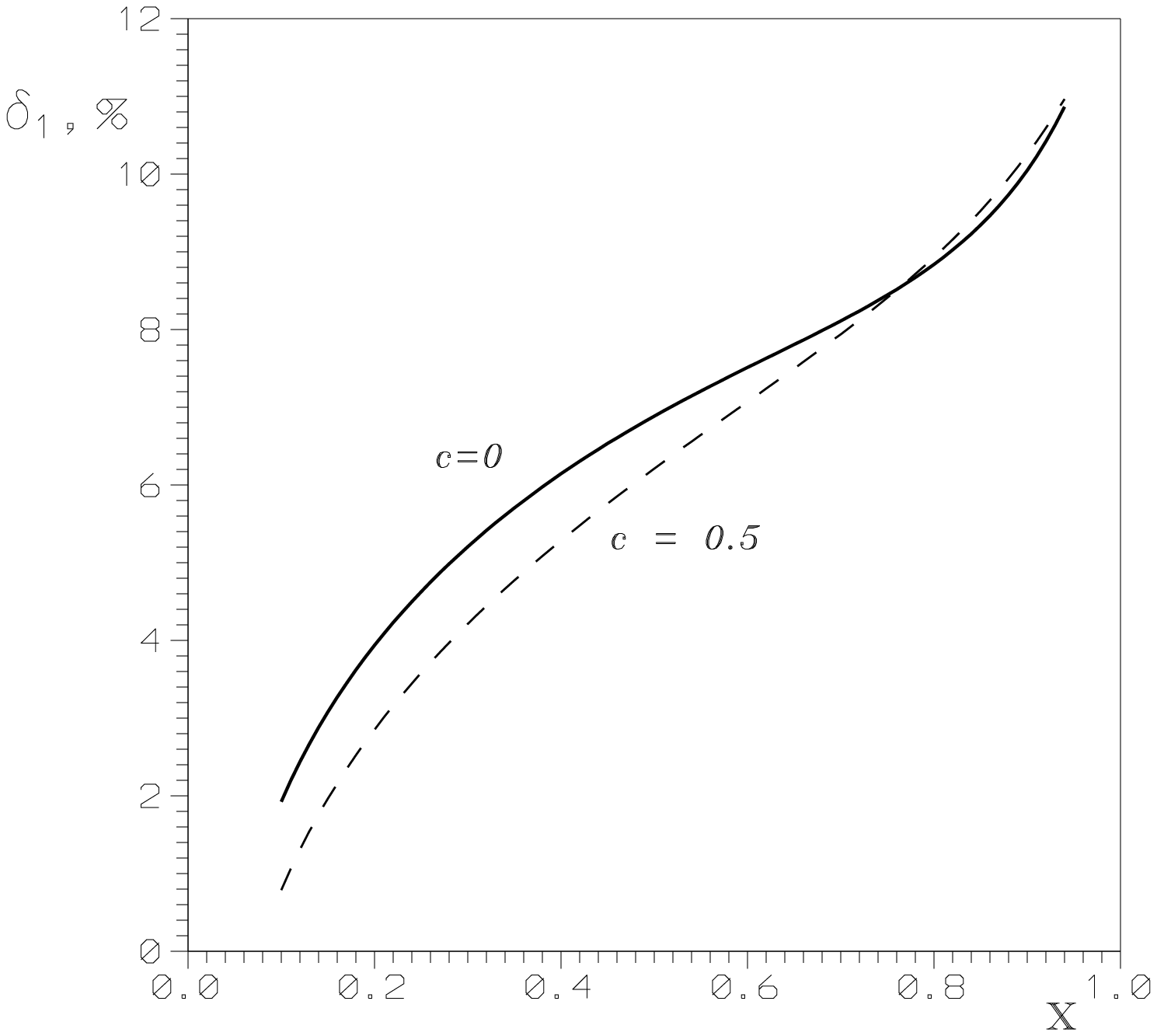,height=10cm,angle=0}}
\end{center}
\end{minipage}
\nopagebreak

\end{document}